\documentclass[numberedappendix,iop,twocolumn]{emulateapj}
\usepackage{amsmath}
\usepackage{amssymb}
\usepackage{color}
\usepackage{graphicx}
\usepackage{natbib}
\usepackage{helvet}
\usepackage{multirow}

\newcommand{\imag}{\ensuremath{i}}
\newcommand{\healpix}{{\sc HealPix}}
\newcommand{\planck}{{\sc Planck}}
\newcommand{\lenspix}{{\sc LensPix}}
\newcommand{\camb}{{\sc CAMB}}
\newcommand{\flints}{{\sc Flints}}

\begin{document}

\title{Fast and optimal CMB lensing using statistical interpolation on the sphere}
\date{\today}
\author{Guilhem Lavaux}
\affiliation{Department of Physics, University of Illinois at Urbana-Champaign, 1002 W Green St, Urbana, IL, 61801, USA}
\author{Benjamin D. Wandelt}
\affiliation{Institut d'Astrophysique de Paris, UMR 7095 CNRS--Universit\'e Pierre et Marie Curie, 98bis bd Arago, F-75014 Paris, FRANCE }
\affiliation{Department of Physics, University of Illinois at Urbana-Champaign, 1002 W Green St, Urbana, IL, 61801, USA}

\begin{abstract}
  We describe a accurate and fast pixel-based statistical method to interpolate fields of arbitrary spin on the
  sphere. We call this method Fast and Lean Interpolation on the Sphere (\flints{}). 
  The method predicts the optimal interpolated values based on the theory of isotropic Gaussian random fields and provides an accurate error estimate at no additional cost. We use this method to compute lensed Cosmic Microwave Background (CMB) maps precisely and quickly, achieving a relative precision 
  of $0.02\%$ at a \healpix{} resolution of $N_\text{side}=$ 4,096, for a bandlimit of $\ell_\text{max}=$ 4,096 in the same time it takes to simulate the original, unlensed CMB map. The method is suitable for efficient, distributed 
  memory parallelization. The power spectra of our lensed maps are accurate to better than 0.5\% at
  $\ell=$ 3,000 for the 
  temperature, the $E$ and $B$ mode of the polarization. As expected theoretically, we demonstrate that, on realistic cases, this method is between two to three orders of magnitude more precise than other known interpolation methods for the same computational cost.
\end{abstract}

\maketitle

\section{Introduction}

Gravitational lensing of the Cosmic Microwave Background (CMB) is a unique probe of the distribution of mass in the entire visible Universe. The first signatures of this CMB lensing signal have already been seen in cross-correlation with large scale structure templates \citep{SZD07,HHNS08}. The effect of lensing on the CMB power spectra will provide powerful additional constraints on the physics of the dark sector \citep{AS03,SHK06}. Including reconstruction of the lensing potential in cosmological parameter analysis removes further degeneracies. Once the lensing deflection is mapped with high precision it will likely allow detecting the absolute mass scale of neutrinos \citep{KKS03,LPPP06} and provide tight constraints on the presence of dark energy at redshifts $z\sim2$. Lensing creates polarization $B$ modes by rotating the stronger $E$ modes \citep{ZS97}. These lensing $B$ modes are not affected by cosmic variance and can therefore significantly improve reconstructions of the lensing potential \citep{SCR06}. The modeling of lensing $B$ modes is further motivated since they are a foreground for the search for the $B$-mode signal of inflationary graviational waves \citep{KS02,KCK02}.

\planck{} will yield the first all-sky temperature and polarization maps
of the CMB with sufficient signal to
noise to detect this signal.  Ground based
experiments such as QUBIC \citep{QUBIC}, SPTpol \citep{SPTpol} or  QUIET \citep{QUIET} contain detailed information about
the fine-scale structure imprinted by lensing while still covering
significant portions of the sky. 
For the extraction of lensing and polarization science from these data sets, fast and precise methods to simulated lensed all-sky CMB maps are indispensable.

In the Born
approximation, which specifies that the lensing effects may be
modeled by the impact of a single lens plane on the CMB,
the lensing corresponds to looking up at the value of the temperature
or the polarization at displaced position on the sky. Thus, making
lensed maps is effectively a resampling of the CMB on a different set
of positions. Different technical solutions were proposed. Currently
three main techniques are known.

The first technique consists in doing a brute force resummation of the
spherical harmonics at a new set of positions. Acceleration methods using symmetry properties of the sampling positions can not be used. While prohibitively expensive for practical use, this method is exact and we therefore employ it on a subset of pixels as a precision benchmark for other methods.

The second technique is polynomial interpolation on
the sphere to compute the value of the temperature and polarization
field at the displaced positions. The most simple of the interpolation technique is a bilinear interpolation of fields on the sphere (hereafter called the naive technique). Another algorithm, a Lagrange polynomial interpolation on the Equi-cylindrical Projection, was first proposed by \cite{HPSSB04} and then later used by \cite{DB08} for a full-sky simulation of the CMB lensing on a light cone of a cosmological simulation. A variant of this method, involving a bicubic interpolation scheme, is implemented in
the publicly available code
\lenspix\footnote{\url{http://cosmologist.info/lenspix}} described in
\cite{Lewis05} and \cite{HL08} (hereafter the {\sc ECP} algorithm).

A third technique has been recently developed by \cite{Basak08}. This algorithm 
consists in recasting the spherical harmonic coefficient of the
unlensed field into the Fourier basis in the $(\theta,\phi)$
variables. Then, we may use a Non-equispaced Fast Fourier Transform
(NFFT) to compute the field at the displaced positions. This method achieves 
high precision and is significantly faster than the first
technique though still too slow to allow for the production of
a large set of lensed maps, which is required for the statistical analysis of observed
CMB data.

We propose a fourth technique that relies on the statistical
properties of the considered fields to be lensed. It is based on the
idea of interpolating the original field but using the known spectral
information to compute the correct weighing coefficients for the
interpolation. This method is related to a Wiener filter
\citep{Wiener49} but not limited to pure Gaussian random field.
We call this method Fast and Lean Interpolation
on the Sphere (\flints{}).\footnote{The reference implementation is written in C++/OpenMP
  and is available at
  \url{http://www.iap.fr/users/lavaux/flints.php}.} In
  this work, we analyze the result obtained based on a \healpix{}
  pixelization. We note that our framework
  may be used on any pixelization of the
  sphere, including the Equi-cylindrical projection already used in previous works. The implementation that we propose takes advantage of the geometrical properties of \healpix{} for a number of memory and computational optimizations.

We note that our interpolation method is of general interest beyond its application to lensing. In fact, since it is based on Wiener filtering it is guaranteed to give the best possible mean squared error of any method for a field with the same power spectrum and for fixed interpolation stencil.

The structure of this paper is the following. In
Section~\ref{sec:interpolating}, we describe the general interpolation
method. Then, in Section~\ref{sec:applens}, we discuss its performance
in generating lensed CMB maps. In Section~\ref{sec:conclusion}, we
conclude.

\section{Interpolating CMB fluctuations}
\label{sec:interpolating}

We propose a direct, simple though sufficiently precise method of
interpolating complex fields on the sphere. This method is based on
the most likely value an isotropic Gaussian random field takes at an arbitrary location
on the sphere, given a set of sampled values (the interpolation stencil) and its power spectrum $C_l$. In Section~\ref{sec:spin0}, we
study the general method for the special case of spin-0 fields. Then,
we generalize the obtained equations to spin-$s$ fields in
Section~\ref{sec:spins}. We then detail in Section~\ref{sec:ngb} the
algorithmic steps used to compute the identifiers of the pixels that are used
in the interpolation. In Section~\ref{sec:tableS}, we describe the
memory and computational time optimizations that we used in \flints{}. We then list all the steps required to achieve the interpolation in Section~\ref{sec:i_algorithm}.
Finally, in Section~\ref{sec:rawperf}, we discuss the results of the raw
performance tests of our interpolation method.

\subsection{Interpolating a scalar field}
\label{sec:spin0}

\begin{figure}
  \includegraphics[width=\hsize]{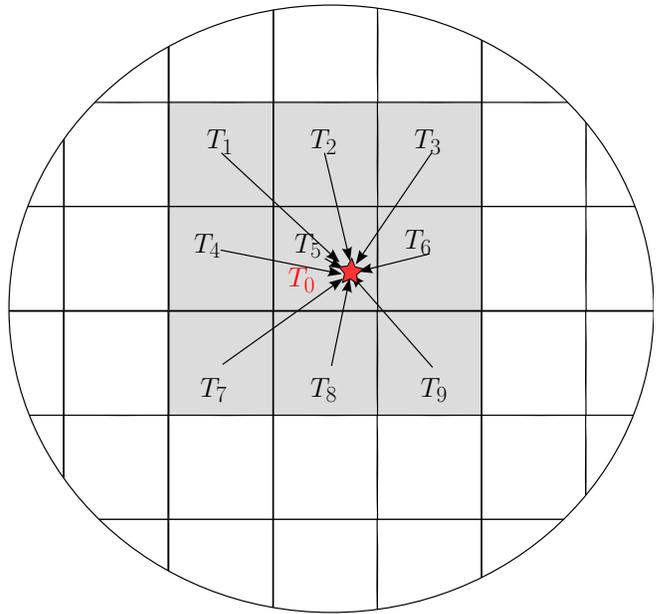}
  \caption{\label{fig:interpolator_picture} {\it Illustration of our interpolation procedure} -- We present here an illustration of the procedure. To get the value of the field interpolated at the position illustrated by the red star, we use the value of this field sampled on the grid given by the squares. In this setting only the nine nearest neighbors, with temperature value $T_i$, contribute to the value $T_0$.}
\end{figure}

We start by considering a Gaussian random scalar field on the sphere
given by the function $T(\hat{n})$, in the direction $\hat{n}$. The
general form of the joint probability of the value of $T(\hat{n})$ in
$N+1$ directions is:
\begin{multline}
  \mathcal{P}\left(T(\hat{n}_0),T(\hat{n}_1),\ldots,T(\hat{n}_N)\right) = \\
     \frac{\sqrt{|\det W|}}{(2\pi)^{(N+1)/2}} \text{exp}\left(-\frac{1}{2\sigma_0^2}\sum_{i,j=0}^{N}
W_{i,j} T(\hat{n}_{i}) T(\hat{n}_{j}) \right).
\end{multline}
with $W$ the inverse of the correlation matrix being
\begin{equation}
  W = V^{-1},
\end{equation}
with $V_{i,j}$ defined for two directions $\hat{n}_i$ and $\hat{n}_j$ as
\begin{equation}
  V_{i,j} = \frac{1}{\sigma_0^2} \zeta(\hat{n}_i \cdot \hat{n}_j), \label{eq:fundamental_variance}
\end{equation}
with the correlation function $\zeta(\cos \theta)$ for two directions separated by an angle $\theta$
\begin{equation}
  \zeta(\cos \theta) = \sum_{\ell=1}^{\ell_{max}} \left(\frac{2\ell+1}{4\pi}\right) C_\ell P_\ell(\cos \theta), \label{eq:correlation_func}
\end{equation}
and the intrinsic variance of the temperature field
\begin{equation}
  \sigma_0^2 = \zeta(\cos(0)=1).
\end{equation}
Now, we can compute the conditional probability of the value of $T(\hat{n}_0)$ given the $N$ other
values:
\begin{multline}
  \mathcal{P}\left(T(\hat{n}_0)|T(\hat{n}_1),\ldots,T(\hat{n}_N)\right) = \\
  \sqrt{\frac{W_{0,0}}{2\pi}} \text{exp}\left(-\frac{W_{0,0}}{2}(T(\hat{n}_0)-\bar{T})^2 \right),
\end{multline}
with
\begin{equation}
  \bar{T} = W_{0,0}^{-1} \sum_{i=1}^N W_{0,i} T(\hat{n}_i), \label{eq:best_temperature}
\end{equation}
and $W_{0,0}$ the top-left most element of the $W$ matrix.  So, given
the values in the directions $\{\hat{n}_1,\ldots,\hat{n}_N\}$, we
define the interpolated value in the direction $\hat{n}_0$ to be equal
to the most likely value $\bar{T}$ as defined above. The precision of
the interpolation is given by the amount of allowed fluctuation
$1/\sqrt{W_{0,0}}$. The advantage of this procedure is that
it is flexible in the number of points we take into account
for the interpolation.  Additionally, it remains purely local and the
complexity scales as $O(N^2 \times N_\text{pix})$, with $N$ the number
of of points to compute the interpolation and $N_\text{pix}$ the
number of pixels in the map. This locality allows us to take full
advantage of the parallelism offered by current multi-core CPUs and by distributed computing environments.
An illustration of our interpolation procedure is given
Fig.~\ref{fig:interpolator_picture} for the case of interpolation stencil with 9 elements ("neighbours").

We further increase the speed of the interpolation by precomputing the
covariance matrix linked to the pixelization, that is the inverse of
the matrix $S_{i,j}$ defined by:
\begin{equation}
  S_{i,j} = \frac{1}{\sigma_0^2}\langle T(\hat{n}_i) T(\hat{n}_j) \rangle
\end{equation}
for $\hat{n}_i$ the direction corresponding to the center of a pixel of the map
to be interpolated. Strictly speaking, we would need to recompute these matrices if the angular power spectrum of temperature fluctuations changes. But as the weights are continuous functions of the angular power spectra, two relatively similar spectra should give the same weights. This alleviates the need of recomputing these weights for any single change of the angular power spectra, at the potential cost of a small loss of precision. $S_{i,j}$ is in practice a $9\times 9$ matrix as we have nine
neighbors for any direction of interpolation. Once we have this matrix it is
fast to determine the $W_{0,i}$ by doing block matrix computation
\citep{numericalRecipes}. First, we write the
shape of the $V$ matrix:
\begin{equation}
  V = \left( 
    \begin{array}{cc}
      1 & B^{T} \\
      B & S 
      \end{array}\right). \label{eq:variance_matrix}
\end{equation}
We let the block matrix shape of its inverse $W$ have the following shape
\begin{equation}
  W = \left(  \begin{array}{cc}
    1/\widetilde{\sigma}_0^2 & \widetilde{B}^{T} \\
    \widetilde{B} & \widetilde{S} 
    \end{array}\right)
\end{equation}
and write that the product should make identity:
\begin{equation}
  V\times W =
  \left( 
    \begin{array}{cc}
      \displaystyle \left(\frac{1}{\widetilde{\sigma}_0^2} + B^{T} \widetilde{B}\right) & \left(\widetilde{B}^{T} +
B^{T} \widetilde{S}\right) \\[.5cm]
      \displaystyle \left(\frac{B}{\widetilde{\sigma}_0^2} + S \widetilde{B}\right) & \left(B \widetilde{B}^{T} + S \widetilde{S}\right)
    \end{array}
  \right),
\end{equation}
which leads us to the following equalities:
\begin{eqnarray}
  \widetilde{\sigma}_0^2 & = & (1 - B^{\dagger} S^{-1} B), \label{eq:variance_interp} \\
  \widetilde{B} & = & -\frac{1}{\widetilde{\sigma}_0^2}S^{-1} B, \label{eq:correlation_interp} \\
  \widetilde{S} & = & \frac{1}{\widetilde{\sigma}_0^2} S^{-1} B B^{\dagger} S^{-1} \label{eq:notused_interp}.
\end{eqnarray}
The matrix $\widetilde{S}$ gives the correlation between the different
direction of the pixelization, $\widetilde{B}$ gives the correlation
between the sought interpolated direction and the direction of the
pixelization, $\widetilde{\sigma}_0$ is the standard deviation of the interpolator
in the interpolated direction. The equations
(\ref{eq:variance_interp}) and (\ref{eq:correlation_interp}) are
computed as needed for each interpolated direction. The equation
\eqref{eq:notused_interp} is not used as we do not need this part of
the matrix.

Finally, we may express the value of the temperature in the interpolated direction $\hat{n}$:
\begin{equation}
 \bar{T}(\hat{n}) = \sum_{i,j=1}^N S^{-1}_{i,j} \frac{\langle T(\hat{n}) T(\hat{n}_j)\rangle}{\sigma_0^2} T(\hat{n}_i),\label{eq:best_temperature_field}
\end{equation}
for which the Gaussian variance of the error of this estimator is 
\begin{multline}
 \sigma^2_T(\hat{n}) = \\
  \sigma^2_0 \bigg( 1 - \frac{1}{\sigma_0^4} \sum_{i,j=1}^N S^{-1}_{i,j} \langle T(\hat{n}) T(\hat{n}_i)\rangle \langle T(\hat{n}) T(\hat{n}_j)\rangle\bigg).\label{eq:gaussvar}
\end{multline}
The above Eq.~\eqref{eq:best_temperature_field} is the strict equivalent of Eq.~\eqref{eq:best_temperature}. We now have both an interpolated value and an estimate of the interpolation error. Even though we used a Gaussian field theory to derive these estimators, the interpolation is also optimal in a least-squared sense for non-Gaussian field.

The time complexity of a \healpix{} spherical harmonic transform \citep{HEALPIX} is  $O(N_\text{side} \ell_\text{max}^2)$, with $\ell_\text{max}$ the number of $\ell$ modes used for the spherical harmonic transform. The difference in complexity between the
\healpix{} spherical harmonic transform and our method means that if the
number of neighbors $N$ is sufficiently small then our method would
be faster to compute a complete CMB sky in any direction than
generating a very high resolution \healpix{} map. This improvement may
be of order $O(\ell_\text{max}^2/(N^{2} N_\text{side,high}))$ for large $\ell_\text{max}$, where $N_\text{side,high}$ is the high resolution map
needed for the naive technique.

 We note that, in the high resolution regime, the scaling
   of this algorithm is like the one of the {\sc ECP} technique
   \citep{HPSSB04,HL08}. On the other hand, there are several
   advantages to our procedure. First, the method provides an accurate error estimate for no additional computational cost.
   The weights take into
   account the part of the signal which is not well represented by the
   pixelization. The ECP interpolation implicitly assume that all the
   information is stored completely in a neighborhood of the direction
   of interpolation, whereas our interpolation takes into account the
   possible lack of complete information. Second, {\flints{}} does
   not need another specific spherical harmonic transform to generate
   the lensed fluctuation on the sphere. It may use a previously
   existing or independently generated map.

\subsection{Interpolating spin-$s$ fields}
\label{sec:spins}

We may now adapt our method for the case of spin-$s$ fields on the
sphere. The concepts are the same, except that we must now handle
complex fields.  We want to interpolate the complex field
$P(\hat{n})$. We assume that this field is isotropic and its
correlation function is defined by:
\begin{eqnarray}
  \langle P^*(\hat{n}') P^(\hat{n})\rangle & = & \sum_{l=s}^{+\infty} \sqrt{\frac{2\ell+1}{4\pi}} C_{P,l} \left({}_s{}Y_{l,-s}(\beta,\alpha)\right) \text{e}^{-i s \gamma}, \\
   & = & \sum_{l=s}^{+\infty} \left(\frac{2\ell+1}{4\pi}\right) C_{P,l} d^\ell_{-s,-s}(\beta) \text{e}^{-i s (\gamma +\phi)},\\
  & = & _{s}\zeta(\alpha,\beta,\gamma),
\label{eq:spins_correl}
\end{eqnarray}
with $\alpha$,$\beta$,$\gamma$ the Euler angles defining the rotation to
transform $\hat{n}'$ into $\hat{n}$ and $_{s}Y_{l,m}$ the spin $s$ spherical harmonic (see Appendix~\ref{app:spin_weighted_Ylm} for the definition).

\begin{figure}
  \includegraphics[width=\hsize]{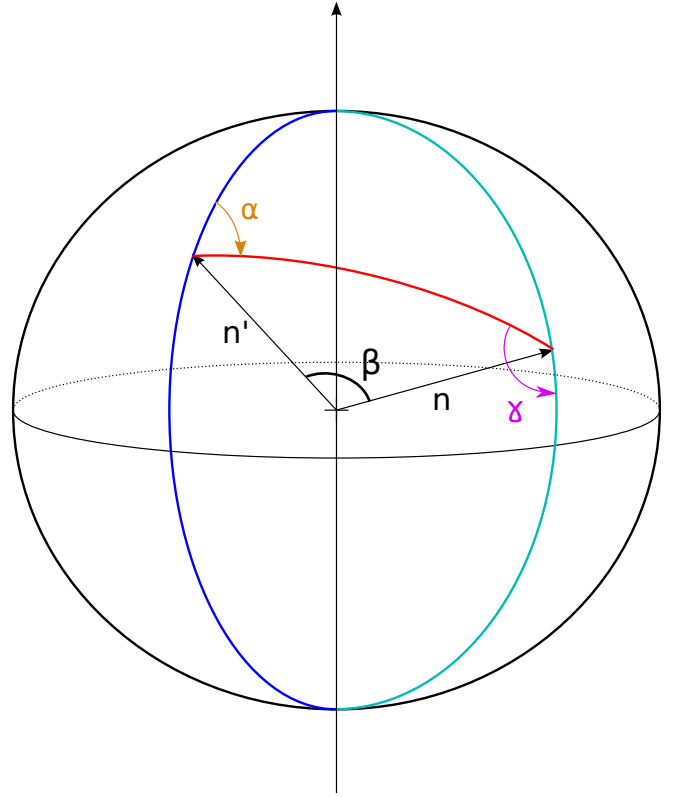}
  \caption{\label{fig:euler_angles} {\it Euler angle convention} -- We
represent here the conventions used for the orientation and the value of the
Euler angles used in Eq.~\eqref{eq:spins_correl}.}
\end{figure}

The function
$d_{-s,-s}^l$ is the Wigner $d$ function. An algorithm to compute this function is recalled in Appendix~\ref{app:wignerd}, and has already been detailed in 
\cite{Trapani06}.
An illustration that defines more properly these angles is
given in Fig.~\ref{fig:euler_angles}.

 As in Eq.~\eqref{eq:fundamental_variance}, we define the Hermitian correlation matrix for two directions $\hat{n}_i$ and $\hat{n}_j$, $0 \leq i,j \leq N$:
\begin{equation}
  V_{i,j} = \frac{{}_{s}\zeta(\alpha_{i,j},\beta_{i,j},\gamma_{i,j})}{{}_{s}\zeta(0,0,0)},\label{eq:v_spins}
\end{equation}
with $\alpha_{i,j}$, $\beta_{i,j}$ and $\gamma_{i,j}$ the Euler angles
defining the rotation to transform $\hat{n}_i$ in
$\hat{n}_j$. Similarly as in Eq.~\eqref{eq:variance_matrix}, we define
the Hermitian matrix $S$ corresponding to the correlation of the
values taken by the \healpix{} pixels, which corresponds here to the
indices $1 \leq i,j \leq N$.

Using some geometry on the sphere it is possible to compute the above
angles.  The relation between $\hat{n}'(\theta',\phi')$,
$\hat{n}(\theta,\phi)$ and
$(\alpha,\beta,\gamma)$ is given by:
\begin{align}
  \cos\beta & = \hat{n}.\hat{n}', \\
  \beta & > 0, \\
  \cos\alpha & = - \frac{\left(\sin\theta \cos\theta' \cos(\phi-\phi') - \cos\theta \cos\theta' \right)}{\sin\beta}, \\
  \sin\alpha & = \frac{\sin\theta}{\sin \beta} \sin(\phi'-\phi), \\
  \cos\gamma & = - \frac{\left(\sin\theta' \cos\theta \cos(\phi-\phi') - \cos\theta'\cos\theta \right)}{\sin\beta}, \\
  \sin\gamma & = -\frac{\sin\theta'}{\sin\beta} \sin(\phi'-\phi).
\end{align}
The choice of the sign of $\beta$ is arbitrary, but the sign of the other angles
depend on this original choice. 
As we need to compute the inverse of the above trigonometric identities we
have decided to use the inverse of the tangent function to improve numerical
stability, taking care of the signs to recover the correct $\alpha$ and $\gamma$
angles in the range $[-\pi;\pi]$.

\begin{table*}
  \begin{center}
    \begin{tabular}{cccccccc}
      \hline
      original $N_\text{side}$ & target $N_\text{side}$ & $\ell$ & Neighbors & Precision & Computational time & Memory & \healpix{} time  \\
            &       &       &    &   & (serial,minutes) & (Megabytes) & (serial,minutes) \\
      \hline
      \hline
      1,024 & 2,048 & 4,096 & 9 & $2\times 10^{-2}$ & 0.8 & 440 & 2.7 \\
      1,024 & 2,048 & 4,096 & 36 & $1\times 10^{-2}$ & 5.1 & 5,500 & 2.7 \\
      2,048 & 4,096 & 4,096 & 9 & $3\times 10^{-3}$ & 3.3 & 1,700 & 5.5  \\
      2,048 & 4,096 & 4,096 & 36 & $3\times 10^{-4}$ & 19 & 22,000 & 5.5 \\
      4,096 & 8,192 & 4,096 & 9 & $4\times 10^{-4}$ & 14 & 6,800 & 12  \\
      \hline
    \end{tabular}
  \end{center}
  \caption{\label{tab:performance} Performance for maps supersampling}
  We give here the computing performance of our method to interpolate a scalar field containing power up to a bandlimit of $\ell =4,096$. The computational time corresponds to the single processor wall time taken by the algorithm to interpolate the map at the given resolution from a map at $N_\text{side}$ to a map at $2\times N_\text{side}$. The \healpix{} computational time is estimated on the spherical harmonic synthesis transform at the target $N_\text{side}$ resolution. The memory consumption gives the size of the pixelization cache in Random Access Memory. The precision corresponds to the square root of the average variance of the error predicted by FLINTS, this square root is divided by the standard deviation of the maps. The quoted serial times have been measured on an Intel Xeon E5410 2.33 GHz based computer. The standard deviation of the maps is 39~$\mu$K/K.
\end{table*}

The probability distribution of the complex field $P(\hat{n}_0)$ given the other value of this field in the direction $\hat{n}_i$, $i=1\ldots N$, is as in Section~\ref{sec:spin0},
\begin{multline}
  \mathcal{P}\left(P(\hat{n}_0)|P(\hat{n}_1),\ldots,P(\hat{n}_N)\right) = \\
  \sqrt{\frac{W_{0,0}}{2\pi}} \text{exp}\left(-\frac{W_{0,0}}{2}|P(\hat{n}_0)-\bar{P}|^2 \right),
\end{multline}
with
\begin{equation}
  \bar{P} = W_{0,0}^{-1} \sum_{i=1}^N W^{*}_{0,i} P(\hat{n}_i).
\end{equation}
The Equations~\eqref{eq:variance_interp},
\eqref{eq:correlation_interp} and \eqref{eq:notused_interp} are still
valid but this time for complex matrices and vectors. In that case
$A^{\dagger}$ is the conjugated transpose of matrix $A$. The resulting
equation for the interpolated value $\tilde{P}(\hat{n})$, taken to be
$\bar{P}$ is
\begin{equation}
  \tilde{P}(\hat{n}) = \sum_{i,j=1}^N S_{i,j}^{-1} \frac{\langle P(\hat{n}) P^{*}(\hat{n}_i) \rangle}{{}_s \zeta(0,0,0)} P(\hat{n}_i),\label{eq:major_eq}
\end{equation}
with $S$ as defined above in this Section. The variance of the interpolated value is
\begin{multline}
   \sigma^2_P(\hat{n}) = {}_s \zeta(0,0,0) \times \\
   \bigg( 1 -  \sum_{i,j=1}^N S^{-1}_{i,j} \frac{\langle P(\hat{n}) P^{*}(\hat{n}_i)\rangle \langle P(\hat{n}_j) P^{*}(\hat{n})\rangle}{{}_s \zeta(0,0,0)^4}\bigg).\label{eq:gaussvar_pol}
\end{multline}
In the following, we focus on the actual implementation of the
interpolation algorithm which uses the above two equations to compute
the interpolated value in any direction.

\subsection{Identifying neighbors}
\label{sec:ngb}

To be fast, the interpolation procedure must rely on a limited number
of pixels, and more particularly the pixels just in the immediate
vicinity of the direction of interpolation, using those pixels which carry most of the information about the interpolated point.

We discuss here a particular implementation to identify neighbors that relies on the \healpix{} framework. 
We use a method that
relies on the use of the \verb,neighbors(), function, which returns
the immediate nine neighbors, sorted geometrically, see 
Fig.~\ref{fig:interpolator_picture}. The time complexity of
\verb,neighbors(), is constant both in the \verb,NESTED, or
\verb,RING, mode of \healpix{}. 

A fast way of extending this algorithm to a higher number of
neighbors consists in using the \verb,NESTED, mode of \healpix{}. This
extension has already been described in \cite{Wandelt98}.
We only consider here the neighbors symmetric according to the
direction in which to interpolate in the \verb,NESTED, tree sense. The
neighbors at a level $y \ge 1$ in a map at resolution $N_\text{side}$
can be found using a pixelization at $N_\text{side}/y$. Their number
is then $9\times 4^{y-1}$. We derive their identifiers in
\verb,NESTED, mode by computing the first neighbors using the function
\verb,neighbors(), of \healpix{} at resolution
$N_\text{side}/y$. We shift the bits of these identifiers by $2
(y-1)$ and fill up the lower bits with all the possible
combinations. This procedure yields all the $9\times 4^{y-1}$ pixel
neighbors of a given direction.

This procedure is more attractive given our computational constraints
than using the alternative procedure \verb,query_disc(),: it allows a stable number of pixels per
neighbors, their geometrical distribution according to the central
direction is stable and insensitive to numerical rounding errors, which makes it easier to tabulate $S^{-1}$ and it is relatively fast to compute the list. For these reasons, we only use this procedure in the
rest of this work.

\begin{figure*}
  \begin{center}
    \includegraphics*[width=.7\hsize]{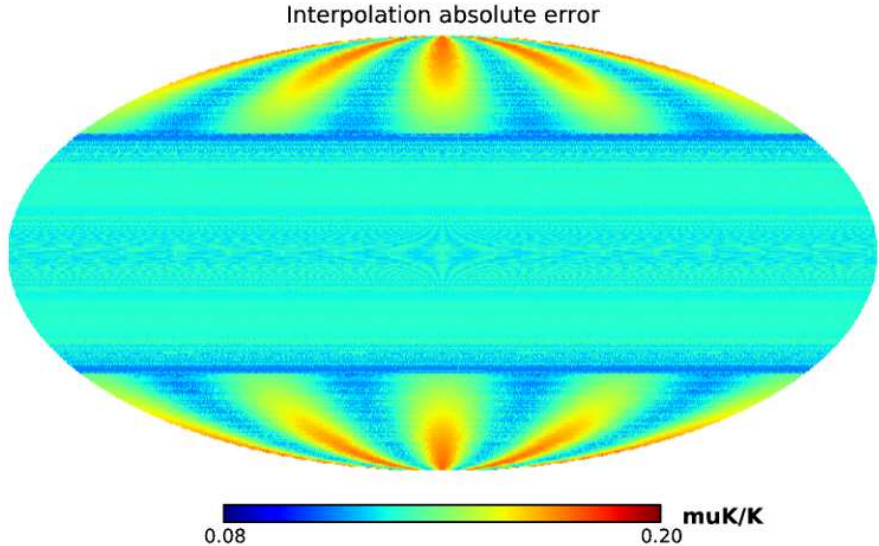}
  \end{center}
  \caption{\label{fig:error_prediction} {\it Spatial distribution of predicted interpolation error for the upsampling} -- We show the distribution of the error predicted by the algorithm for each interpolated pixel of the sky. We predict a map at $N_\text{side}=4,096$ from a map at $ N_\text{side}/2 =$ 2,048, with $\ell_\text{max}=4,096$. The structure of the error distribution reflects the features of the \healpix{} pixelization. }
\end{figure*}

\subsection{Tabulating $S^{-1}$}
\label{sec:tableS}

To reduce the time needed to compute the interpolation, we tabulate
the matrices $S^{-1}$ for all possible central pixels (corresponding to
$T_5$ in the Figure~\ref{fig:interpolator_picture}). This means we
would need to precompute and store a matrix for each pixel of the map
to interpolate. We made use of the symmetries of the \healpix{}
pixelization to reduce this cost.  As there is a rotational  invariance in the equatorial part of the pixelization, 
it suffices to store one pixel per ring. For the polar regions, it would be sufficient to store the weights for only one quarter of one of  the  \healpix{} base tiles, since there is an eight-fold rotational invariance and an additional north-south symmetry. Using all these symmetries would reduce the amount of memory required to store the precomputed weights by nearly $1/48$.
 
For our implementation we use a subset of these symmetries since there is a trade-off between reduction of memory use and the implementation complexity for the required pixel permutations. In addition to the equatorial symmery, we
use the north-south symmetry for the case of the spin-0 field only. 
Memory use scales as $O(N_\text{ngb}^2\times
N_\text{side}^2)$, with $N_\text{ngb}$ the number of
neighbors chosen for computing the interpolation.

To compute $S^{-1}$ itself from $S$, we use the Cholesky decomposition
of $S= {}^{t}L L$ and do a direct inversion of $L$. We used the
algorithm described in \cite{numericalRecipes}. As the matrix $S$
becomes so ill-conditioned that the Cholesky decomposition fail for
high resolution map, we introduce an additional term $n$ to help at
computing this decomposition when this term is required. So, in
practice, we decompose $S + n \mathcal{I}$ instead of $S$. By
construction, the value of $n$ is much smaller than one. If $n$ is too
big, the interpolation error is larger than it should. If $n$ is too
small, the decomposition fails. We use for $n$ a value given by
\begin{equation}
  n = -\lambda_\text{min,negative} + \epsilon_\text{precision}
\end{equation}
where $\lambda_\text{min,negative}$ is the lowest negative eigenvalue of $S$ at the used precision, $\epsilon_\text{precision}$ is a quantity dependent on the machine precision used for doing the actual computation on $S$. For double-precision, we take $1.49\,10^{-8}$, which is the square root of the smallest deviation from $1.0$ detectable in a double precision representation.

\subsection{The interpolation algorithm}
\label{sec:i_algorithm}

Finally, the interpolation algorithm consists in achieving the following steps:
\begin{enumerate}
\item We set the resolution of the map to interpolate at $N_\text{side}$ and the band width to $\ell_\text{max}$.
\item We start by tabulating according to $\beta \in [0,\pi]$ the $\zeta(0,\beta,0)$ function, defined in Eq.~\eqref{eq:spins_correl}.
\item For the \healpix{} pixelization at $N_\text{side}$, we compute $S$ using Eq.~\eqref{eq:v_spins}, invert it and tabulate it according to Section~\ref{sec:tableS}.  This is the end of the preparation phase.
\item For any direction $(\theta,\phi)$, we compute the identifiers of the pixel neighbors according to Section~\ref{sec:ngb}.
\item We sum up the value of the field sampled at those pixels and weighed them according to Eq.~\eqref{eq:major_eq}.
\item If required by the user, we also compute the error in the interpolation using Eq.~\eqref{eq:gaussvar_pol}.
\end{enumerate}
In all the subsequent tests, we have used the above scheme to generate the interpolated values.

\subsection{Supersampling interpolation performance}
\label{sec:rawperf}

We test the interpolator by 
supersampling a CMB map to twice the original resolution.
Please refer to the results in
Table~\ref{tab:performance} and in
Figure~\ref{fig:error_prediction}. In all these tests, we start from a
map at a resolution of $N_\text{side}$ and predict its values at twice the resolution, $2\times N_\text{side}$.  Note that all these test are done for fixed bandlimit $\ell=4,096$ while the resolution of the grid is varied. For the lowest starting resolution of $N_\text{side}=1,024$ the field is somewhat undersampled, since modes beyond $\ell=2N_\text{side}$ begin to be noticeably aliased on the \healpix{} grid.
Owing to the hierarchical property of the \healpix{} grid, the higher resolution pixels tile the lower resolution pixels. The distance from the members of the interpolation stencil is therefore one fourth or three fourth of the size of a pixel. Based on numerical experiments we estimate the worst case error for directions farthest away from the members of the interpolation stencil to be no more than 50\% higher than shown in Figure~\ref{fig:error_prediction}.

In Figure~\ref{fig:error_prediction}, the distribution of the
interpolation error reflects the features of the \healpix{}
pixelization. The error is uniformly small in the equatorial regime and shows the symmetries of the pixelization in the northern and southern caps.

We tabulate the supersampling precision and the time consumption in Table~\ref{tab:performance}. For comparison we list the time required to compute a
single \healpix{} transform from $\{a_{lm}\}$ space to
pixel space. The two operations take roughly the
same time. Note that the spherical harmonic transform has only
been done one way, whereas the interpolation starts from
pixels and yields pixels. As expected the error decreases with increased
resolution and number of neighbors. It is
interesting to see that we may get a decrease of one
magnitude in the error by changing the number of neighbors at the
resolution $N_\text{side}=$ 2,048. Doing the same exercise
with $N_\text{side}=$ 4,096 yields a large number of degeneracies in the
$S$ matrices. That shows that we reach the level where the problem of
interpolation is dominated by errors in the floaing point representation, as expected since
a \healpix{} map at resolution $N_\text{side}$ is able to encode wavenumbers up
to $\ell \sim  2N_\text{side}$. So, as our $\ell_\text{max}$ is 4,096
here, a map at $N_\text{side}=$ 2,048 can be
interpolated at high precision with a sufficient number of
neighbors.

\section{Application to CMB lensing}
\label{sec:applications}
\label{sec:applens}

\begin{table*}
  \begin{center}
    \begin{tabular}{c | ccccc | c | cccc}
      \hline
      \multirow{4}{*}{$N_\text{side}$} & \multicolumn{5}{c|}{ {\flints{}}} &  {\sc Torus} & \multicolumn{4}{c}{{\sc ECP}(\lenspix{})}   \\[.2cm]
      & Init. & Interp. & Total & \multirow{2}{*}{Precision} & \multirow{2}{*}{Precision} & Total &  Interp. & Total & \multirow{2}{*}{Precision} & \multirow{2}{*}{Precision} \\
      & time & time & time & $L_2$ & $L_\infty$ & time & time & time & $L_2$ & $L_\infty$ \\
       & (minutes) &  (minutes) & (minutes) & & & (minutes) & (minutes) & (minutes) & \\
      \hline
      \hline
      1,024 & 8 &   1 & 8 & 2\% & 11\% & 32 & 5.0 & 9 & 5\% & 24\% \\
      2,048 & 9 &  6 & 20 & 0.3\% & 2.65\% & $\sim$160-250$^*$ & 12 & 20 &  $9\%$ & 125\% \\
      4,096 & 16 & 22 & 51 & 0.04\% & 0.3\% & $\sim$640-1,000$^{*}$ & 34 & 52 & $7.5\%$ & 127\% \\
      \hline
  \end{tabular}
  \end{center}
  \caption{\label{tab:lensing_perf} Lensing performance}
  NB: Numbers with $^{*}$ are extrapolated from table 3 of \cite{Basak08}. ``Interp.'' stands for ``Interpolation''. ``Init.'' stands for ``Initialization''. We measure the time for producing a lensed map, temperature and polarization, from a random realization of the CMB fluctuations and of the lensing potential. The total time is the sum of the interpolation time, the time to make an unlensed healpix map, the time to compute the deflection map but not the initialization time given in the first column. The  $L_2$ ($L_\infty$) precision corresponds to the standard deviation (maximum absolute value) of the error distribution divided by the standard deviation of the simulated CMB map, which is $39\,\mu$K/K. For all cases we used $\ell_\text{max}=4,096$ and nine neighbors for FLINTS, and a interpolation factor of one for the ECP method in \lenspix{}. The maps at $N_\text{side}=1,024$ were therefore undersampled.
\end{table*}

\begin{figure}[t]
  \includegraphics[width=\hsize]{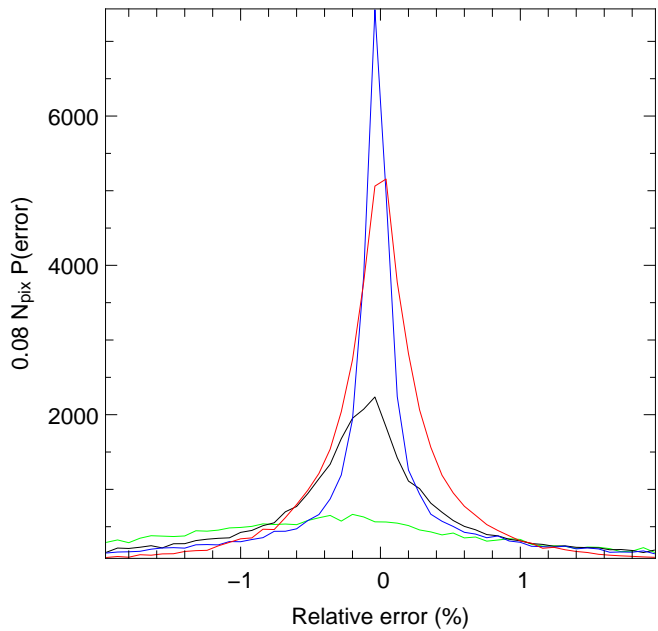}
  \caption{\label{fig:cmb_error}  {\it Interpolated CMB vs True value} --
Distribution of the error in the value given by both naive
interpolation on a \healpix{} mesh and our method for different \healpix{} resolution. For each line,
we represented the distribution of the relative difference between the actual
interpolated value and the true value. The green, black and blue lines correspond to
\healpix{} interpolation at $N_\text{side} =$ 2,048, $N_\text{side} =$ 4,096 and $N_\text{side}=$ 8,192 respectively. The
red line corresponds to our method at $N_\text{side} =$ 2,048.}
\end{figure}

\begin{figure}[t]
 \includegraphics[width=\hsize]{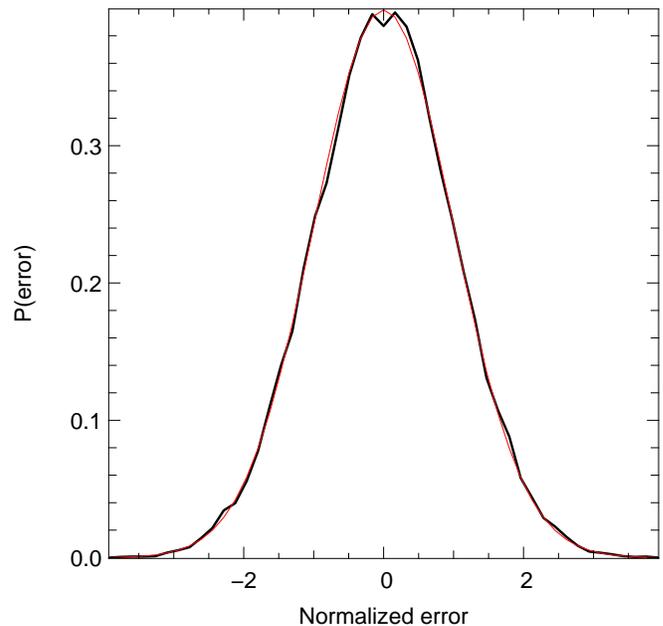}
 \caption{\label{fig:cmb_exp_error} {\it Accuracy of the error estimate} --
 The thick black line shows the measured error normalized using the error estimate, Eq.~\eqref{eq:gaussvar}, at each pixel. The overplotted thin red line is a Gaussian distribution of width $1$ and centered on $0$ showing perfect agreement between predicted and actual error.}
\end{figure}

\begin{figure*}
  \includegraphics[width=\hsize]{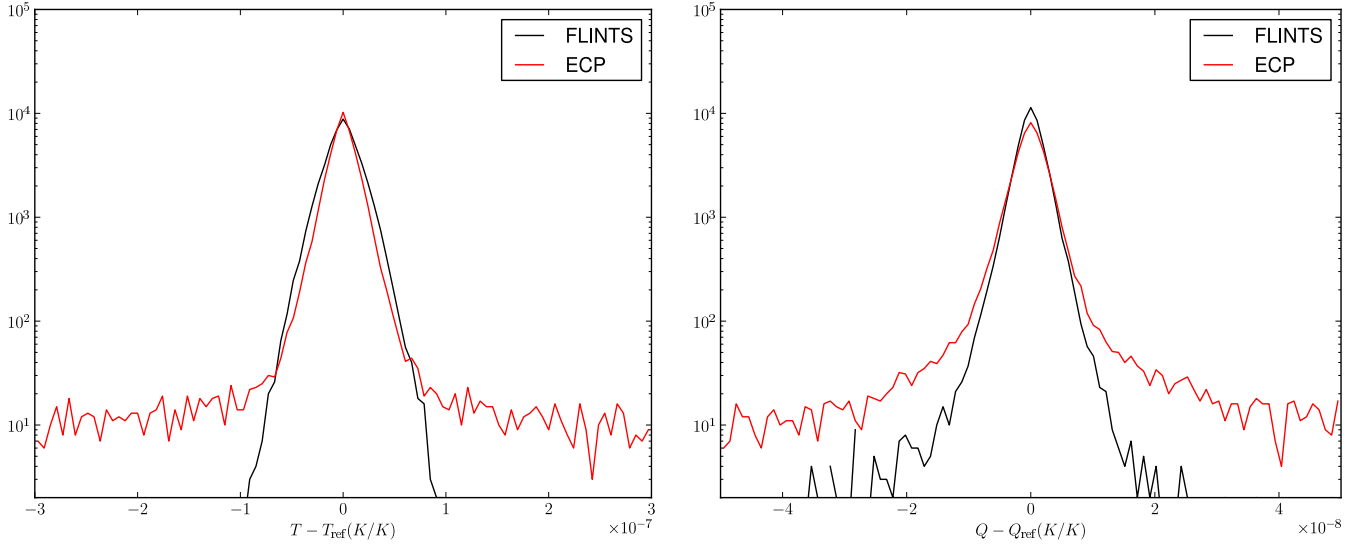}
  \caption{\label{fig:distrib_ecp_flints} {\it Error distribution of the Equicylindrical projection interpolation and of FLINTS} -- We represent the measured error distribution of the temperature ($T$, left panel) and one plane of polarization ($Q$, right panel) map. We use $N_\text{side}=4,096$ and $\ell_\text{max}=4,096$ and  no multiplication factor for the ECP method. The error distribution is computed by taking the difference of the value predicted by the interpolator to the exact value computed using ECP method. In black (red) solid line, we represent the error distribution of our {\flints{}} (ECP interpolation) algorithm.}
\end{figure*}

\begin{figure*}
  \includegraphics[width=\hsize]{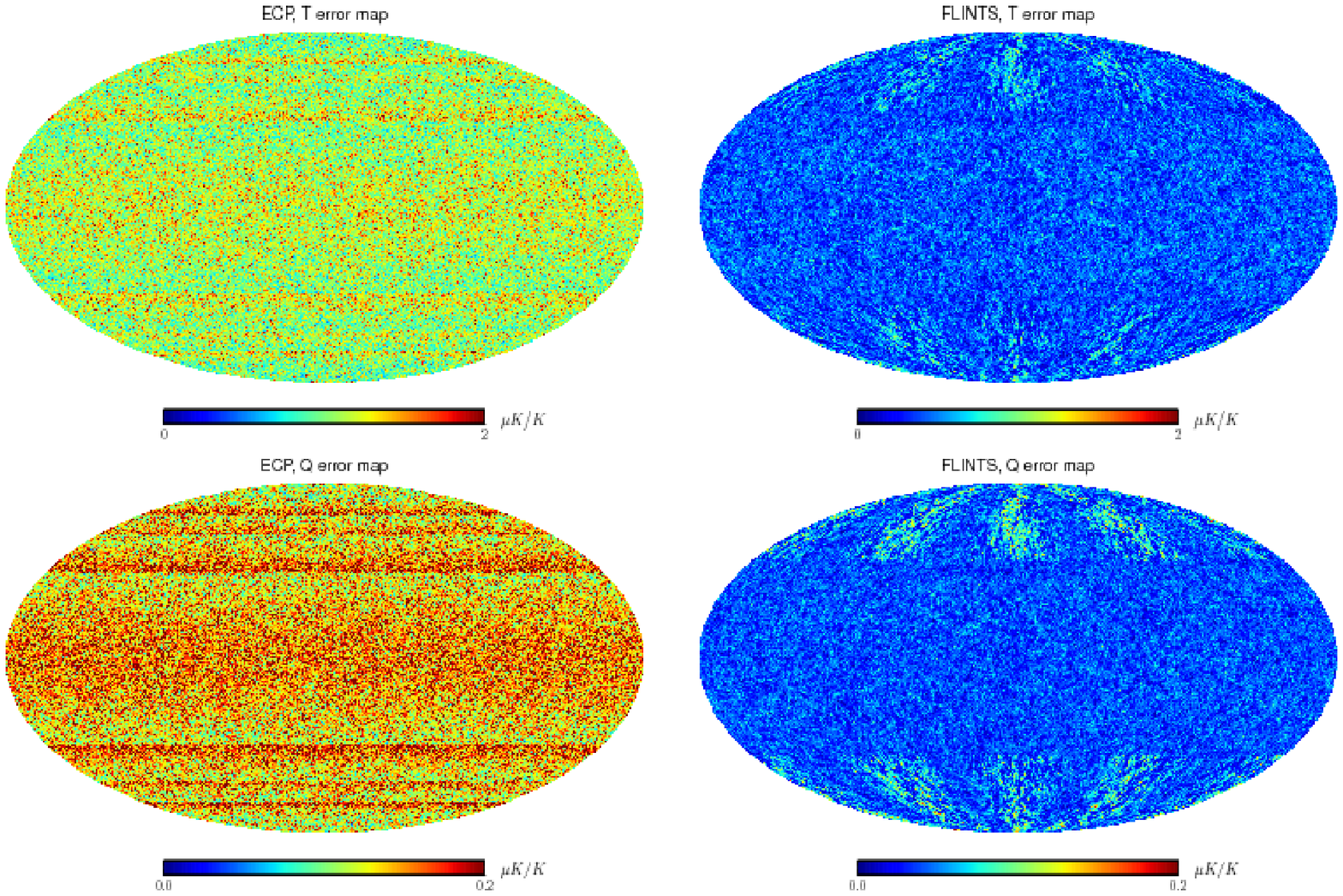}
  \caption{\label{fig:distrib_ecp_flints_sky} {\it Error distribution of the Equi-cylindrical projection interpolation and of FLINTS on the sky} -- We represent a comparison of the error distribution on the sky of the interpolated temperature ($T$, top panels) and one polarization plane ($Q$, bottom panels) map. The errors in the ECP interpolation, are represented in the left panels. The errors of the FLINTS interpolation are represented in the right panels. We compare to the exact lensing method. All maps were computed at $N_\text{side}=1,024$, with $\ell_\text{max}=2,048$. For visualization purposes, we degraded the error maps to a resolution of $N_\text{side}=128$.}
\end{figure*}

In this section, we focus on the use of our interpolation procedure
for producing lensed maps of the CMB.
In Section~\ref{sec:lenstheory}, we recall the basic lensing equation
in the Born approximation and the notation. In
Section~\ref{sec:lens_tests}, we compare the predictive performance of
FLINTS to \lenspix{} and to the naive technique in \healpix{}. Finally, in Section~\ref{sec:lens_ps}, we
compare the power spectra of the lensed temperature and polarization
as obtained through FLINTS and using \camb.

\subsection{Theory of lensing}
\label{sec:lenstheory}

Gravitational lensing acts like a remapping of the CMB
photons on the sky. Thus the temperature signal on the sky is given by
\begin{equation}
  T_\text{observed}(\hat{{\bf n}}) = T_\text{CMB}(\hat{{\bf n}} +
{\bf d}(\hat{\bf{ n}})),
\end{equation}
where $T_\text{observed}$ is the observed temperature of the CMB in the
direction $\hat{\bf n}$, $T_\text{CMB}$ is the unlensed primary signal, and 
${\bf d}(\hat{\bf n})$ is the deflection field. This relation is 
also true for the polarization field defined by
\begin{equation}
 P(\hat{\bf n}) = Q(\hat{\bf n}) + \imag U(\hat{\bf n})
\end{equation}
with $Q$ and $U$ the Stokes parameters, with $P$ being a spin-2 field.

The deflection field defines in
what direction and by what angle the photons were deflected from their original
position. In the local basis defined by the direction $\hat{\bf n}$, we may
define the angle $\alpha(\hat{\bf n})$ as
\begin{equation}
  {\bf d}(\hat{\bf n}) \propto \cos(\alpha) {\bf u}_\theta(\hat{\bf n}) +
\sin(\alpha) {\bf u}_\phi(\hat{\bf n})
\end{equation}
with $(\hat{n},{\bf u}_\theta,{\bf u}_\phi)$ the local spherical orthonormal
basis, with ${\bf u}_\theta = \partial \hat{n}/\partial \theta$ and ${\bf
u}_\phi = \partial \hat{n}/\partial \phi$. In this basis, the lensed direction
$\hat{\bf n}'$ may be written as
\begin{multline}
  \hat{\bf n}' = \hat{\bf n} + {\bf d}(\hat{\bf n}) = \cos(|d|) \hat{\bf n} +
\sin(|d|) \cos(\alpha) {\bf u}_\theta \\
  + \sin(|d|) \sin(\alpha) {\bf u}_\phi.
\end{multline}
We prefer this form of the lensed direction instead of using the angles for
numerical stability, at the cost of computing a few additional trigonometric
function. Using the direct angle relation, as in e.g. \cite{Basak08}, may expose
us to problems in the case of directions near the poles. The displacement field ${\bf d}(\hat{\bf n})$ is obtained by taking the spherical gradient of a scalar potential, corresponding to the projected gravity field in the Born approximation.

\subsection{Test of producing lensed CMB maps}
\label{sec:lens_tests}

We now test our procedure for generating precise lensed CMB maps. We compare our
generated maps to:
\begin{itemize}
\item[-] the true lensed maps, obtained by summing exactly the spherical
harmonics at the position of interpolation
\item[-] the naive interpolation procedure using a simple bilinear interpolation of CMB maps simulated at high
resolution. This algorithm  is included in the \healpix{} package for visualization purposes and was not intended for scientific use, but it is still useful as a point of comparison to assess whether a more complicated interpolation procedure is warranted.
\item[-] the ECP bicubic interpolation algorithm implemented in \lenspix{}. This is the current mainstream algorithm for quickly computing lensed maps.
\end{itemize}

As it is very expensive to compute the true lensed map on the full
sky we limit ourselves to testing our method on a restricted subset of
pixels distributed uniformly over the sky. More specifically we
chose directions on the sky at a resolution of $N_\text{side}=64$
(49152 pixels). 

We show the result in Fig.~\ref{fig:cmb_error}. We
note that \flints{} behaves much better than the naive 
interpolation. Our interpolation procedure,  executed at a resolution
of $N_\text{side}$ is able to match fairly well with the naive
interpolation used at $4 N_\text{side}$. This matches \cite{Lewis05} who indicated that one
needs at least 16 times more pixels than the base CMB map to produce
acceptable spectrum using a naive interpolation procedure.

Moreover, the tails of the error distributions of the interpolated
field are much more Gaussian, as illustrated in
Fig.~\ref{fig:cmb_exp_error}. There, we represented the error
normalized by the expected standard deviation given by
Eq.~\eqref{eq:gaussvar}. If the errors are Gaussian
with exactly this deviation, we must obtain a Gaussian of standard deviation
equal to one, which is exactly what we obtain.

In Table~\ref{tab:lensing_perf}, we compare the performances of our method, {\flints{}}, with the method described in \cite{Basak08}, labelled {\sc Torus}, and the {\sc ECP} interpolation implemented in \lenspix{}. For {\flints{}} and {\sc ECP} we give an estimate of the attained precision. We also measure the time required to produce one lensed map. We note that the precision is better for {\flints{}} than for {\sc ECP}. The problem for {\sc ECP} are the heavy tails in the error distribution, as shown in Fig.~\ref{fig:distrib_ecp_flints}. Furthermore, at low $N_\text{side}$ fixed $\ell_\text{max}$, the field fluctuations are undersampled which degrades the predictive properties of ECP method. On the other hand, {\flints{}} keeps errors lower because it takes into account the underlying fluctuations through the use of the angular power spectra in the weights. At $N_\text{side}=4,096$, the central part of the error distribution represented in Fig.~\ref{fig:distrib_ecp_flints} is the same showing convergence of the two methods.

A comparison of the differences in the spatial distribution of the interpolation errors of the ECP and FLINTS method are given in Fig.~\ref{fig:distrib_ecp_flints_sky}. We represent there the sky distribution of the errors in the lensed temperature and polarization maps. We compute the reference maps using the full resummation of the spherical harmonics at the displaced positions. The maps are computed at a resolution $N_\text{side}=1,024$, $\ell_\text{max}=2,048$, with an oversampling factor equal to one for the ECP method. We note that the error distribution of the ECP interpolation is linked to the projection of the ECP mesh onto an \healpix{} mesh. On the other hand, the FLINTS interpolation is essentially tracking the shape of the \healpix{} grid, as already seen in Fig.~\ref{fig:error_prediction}. As in Fig.~\ref{fig:distrib_ecp_flints}, the overall amplitude of error of the ECP interpolation is larger than the one given by the FLINTS interpolation.

 The total time for both methods is similar but {\flints{}} has initialization time overhead for precomputing $S^{-1}$ (Section~\ref{sec:tableS}).  This overhead may still be further optimized by using more of the symmetries of \healpix{} pixelization. Overall, {\flints{}} is more precise than {\sc ECP} by order of magnitude for simulations resembling data from a high resolution CMB polarization mission.  Both of these methods are much faster than the {\sc Torus} method.

\subsection{Lensed power spectra}
\label{sec:lens_ps}

\begin{figure*}[t]
  \includegraphics[width=\hsize]{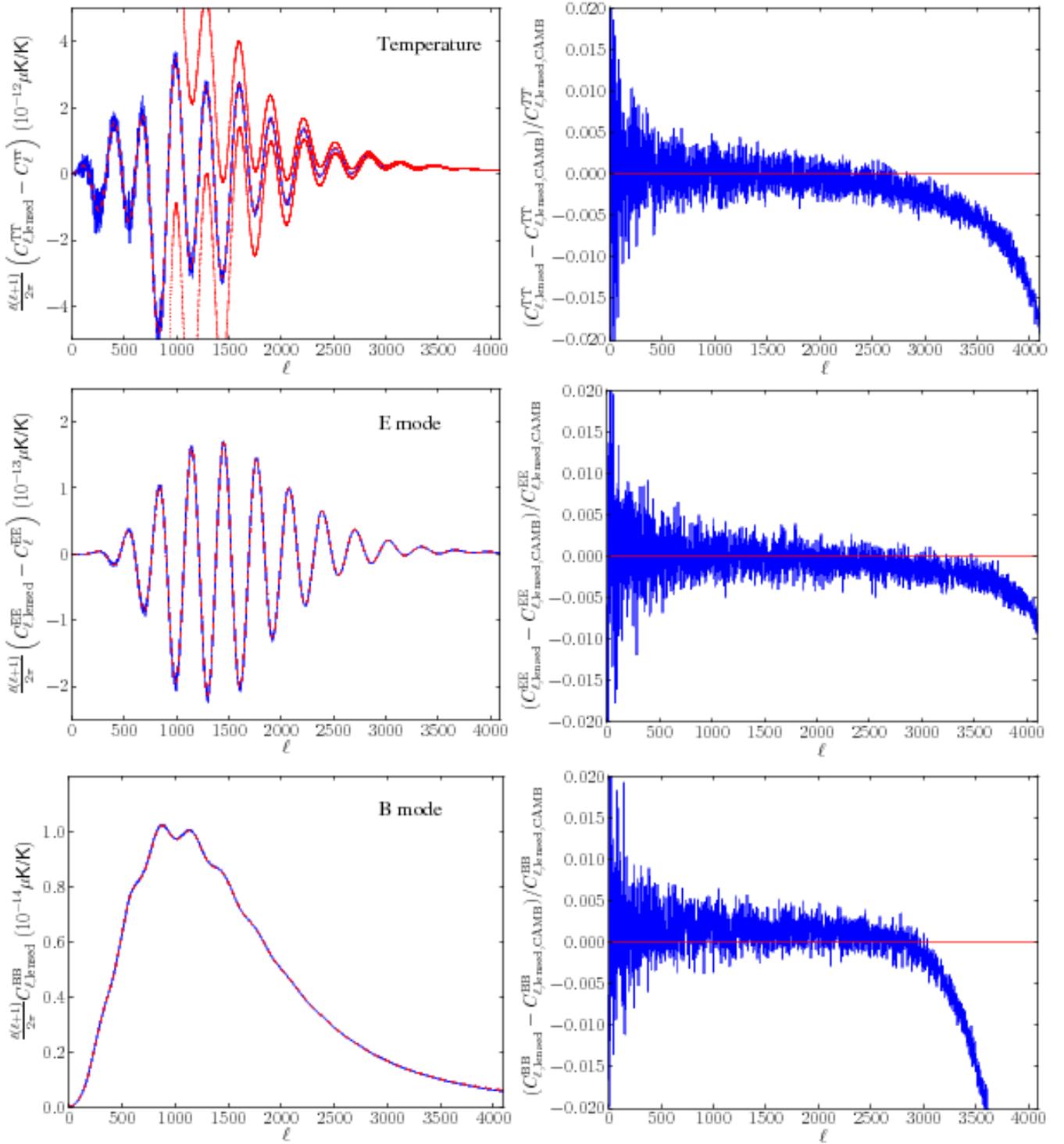}
 \caption{\label{fig:angle_spectra} {\it Precision of lensed power spectra}  -- We
   show the differences between the lensed spectra and the
   unlensed spectra for $\sim$350 realizations of CMB fluctuations and
   lensing potentials. In the left panels,
   the dashed, red line is  the difference as computed by \camb{}, 
   and the solid, blue line the result
   obtained using our interpolation technique. Cosmic variance error bars are shown as a dotted line in the top-left most panel. In the right panels, we represented the relative difference between the \camb{} prediction and FLINTS prediction.
   We represented the
   change in the $C_\ell$ for the temperature (top row), the
   $E$ polarization mode (middle row), the $B$ polarization mode
   (bottom row). 
   The fluctuations are here normalized by the CMB temperature
   and are unitless. The red lines in the right panels represent perfect agreement with \camb{}. The lensed maps were computed using a CMB map at $N_\text{side}=$ 4,096, $\ell_\text{max}=$ 5,000 and $9$ neighbors. } 
\end{figure*}

As an additional test of the precision of our method we compare the
power spectra of our lensed temperature and polarization maps with the
theoretical predictions from \camb{} \citep{CAMB,CL05}.  We show the
difference of the average spectrum of 350 lensed maps and the
unlensed spectrum in Figure~\ref{fig:angle_spectra}.

We tested how our method fared compared to naive
temperature interpolation and on a limited number of exactly
interpolated pixels.  We now check the interpolation of both
the temperature and the polarization fields on different scales
by considering the difference between the lensed spectra, $C^{TT}_{\ell,\text{lensed}}$ and the unlensed spectra.  To assess the precision of our
procedure, we use the spectra computed by \camb{} \citep{CAMB,CL05} as a
reference. The results, given as the difference of the lensed spectra
to the unlensed spectra, are given in
Figure~\ref{fig:angle_spectra}. We used $N_\text{side}=$ 4,096 and
$\ell_\text{max}=$ 5,000 as an input to FLINTS. The \camb{} spectra were
predicted using $\ell_\text{max}=$ 10,000 and $k_{\eta,\text{max}} =$ 100,000 in very high accuracy mode. We also show the relative
difference between our spectra and \camb{} spectra. We see that the
difference between the two does not exceed 0.5\% statistically at
$\ell=$ 3,000. While the B-mode spectrum has a small but systematic excess of power of about $0.1-0.2$\% in our measured BB compared to the prediction given by \camb{}, we note that this is of the same order as the advertised accuracy \camb{} even in high-precision mode \citep{CL05}. In addition, \camb{} is not guaranteed to be accurate at this level beyond $\ell\sim$ 2000, so the comparison breaks down at this point and it is not clear whether \camb{} or \flints{} is more accurate.

In any case, the interpolation accuracy is more than sufficient for practical purposes in all cases. To illustrate, we display the cosmic variance range for the temperature power spectrum in the first panel of Figure~\ref{fig:angle_spectra}. This shows the unavoidable error in the estimation of $C_\ell$ from a perfect, all-sky, unlensed map. For the very small B-mode signal  these numerical errors will remain much smaller than measurement error for the foreseeable future.

\section{Conclusion}
\label{sec:conclusion}

For a given interpolation stencil, we describe the optimal interpolation technique for isotropic band-limited fields of arbitrary spin, 
sampled on the sphere. Taking advantage of the symmetry properties of the 
\healpix{} pixelization, the method is fast and memory-efficient. To test this approach 
we implement a supersampling filter for \healpix{} temperature and polarization maps. 
A Monte Carlo study confirms both the predicted precision and our estimates of memory 
and CPU time scaling.

We demonstrate this interpolation method to be powerful tool to simulate lensed CMB temperature 
and polarization maps from unlensed maps. Our Monte Carlo comparison to exact reference 
maps computed by \lenspix{} and to predicted lensed power spectra by \camb{} demonstrate 
that we achieve an accuracy which exceeds the requirements of the \planck{} data \citep{planckbook} while 
reducing the required computational time by an order of magnitude  compared
to using, e.g., the Torus method.
In addition, the method allows very easy parallelization as the procedure is strictly 
local in pixel space. We compared the performance and the precision of 
our method to the Equi-cylindrical projection interpolation method. The two methods 
have similar speed.  {\flints{}} is more precise and has no catastrophic errors.

We conclude that this method is a very promising
technique in terms of speed, precision and scalability for
the simulation of high resolution maps of the lensed CMB temperature and polarization anisotropies.  
\flints{} enables us to produce lensed maps as cheaply as making a spherical harmonic transform, and makes us capable
of producing thousands of simulations of the lensed sky within an acceptable computational time.
This advance may allow us to run a full likelihood analysis of the lensing potential in observed
CMB data with current computer technology, which is not possible with other known methods of computing lensed
maps.

\section*{Acknowledgments}

We thank the Planck group at JPL and the astrophysics group at Caltech for their hospitality while this work was begun.
We thank S. Prunet for comments on the first version of the draft. We acknowledge financial support from NSF Grant AST 07-08849. 
GL acknowledges grant from the the ``Programme visiteur de l'IAP'' and financial support from French ANR (OTARIE).
This research was supported in part by the National Science Foundation
through TeraGrid resources provided by the NCSA under grant number [TG-MCA04N015]. Teragrid systems are
hosted by Indiana University, LONI, NCAR, NCSA, NICS, ORNL, PSC,
Purdue University, SDSC, TACC and UC/ANL.
The authors thank the anonymous referee for his/her constructive remarks.

\bibliographystyle{apj}

\appendix

\section{Spin-weighted spherical harmonic}
\label{app:spin_weighted_Ylm}

Spin $s$ functions ${}_sf$ transform under a locally planar rotation $R(\theta)$ about the direction $\hat{n}$ as 
\begin{equation}
  {}_s f(R(\theta) \hat{n}) = \text{e}^{i s \theta} {}_s f(\hat{n})
\end{equation}
On the sphere, these functions may be expanded on the spin-$s$ spherical harmonic basis ${}_sY_{\ell,m}(\hat{n})$ as \citep{NP66,GMNRS67,ZS97}
\begin{equation}
  {}_s f(\hat{n}) = \sum_{\ell=0}^{+\infty} \sum_{m=-\ell}^{+\ell} {}_s f_{\ell,m} \left[{}_sY_{\ell,m}(\hat{n})\right].
\end{equation}
As for spin-0 function, the spherical harmonic coefficient ${}_s f_{\ell,m}$ may be obtained using
\begin{equation}
  {}_s f_{\ell,m} = \int_{S^2} \text{d}\Omega(\hat{n}) \left[{}_s Y_{\ell,m}(\hat{n})\right]^{*} {}_s f(\hat{n})
\end{equation}
with $S^2$ being the sphere in dimension three. The spin-weighted spherical harmonic function may be expressed directly from the Wigner rotation matrices: \citep{GMNRS67}
\begin{equation}
  {}_s Y_{\ell,m}(\hat{n}) = {}_s Y_{\ell,m}(\theta,\phi) = \sqrt{\frac{2\ell+1}{4\pi}} D^{\ell}_{-s,m}(\phi,\theta,0)
\end{equation}
where we used the Condon-Shortley phase \citep{CS51} convention, $\theta$ and $\phi$ are respectively colatitude and longitude on the sphere. The Wigner $D$ matrix $D^\ell_{m,m'}$ may be further expanded with the help of the Wigner $d$ function
\begin{equation}
	D^\ell_{m,m'}(\phi,\theta,\rho) = \text{e}^{-i m \phi} d^\ell_{m,m'}(\theta) \text{e}^{-i m' \rho}
\end{equation}
which yields
\begin{equation}
  {}_s Y_{\ell,m}(\hat{n}(\theta,\phi)) = (-1)^s \sqrt{\frac{2\ell+1}{4\pi}} \text{e}^{i m \phi} d^\ell_{m,-s}(\theta).
\end{equation}
Furthermore, we recall the spin-$s$ spherical harmonic addition relation \citep[e.g.][]{HW97}
\begin{eqnarray}
   \sum_{m=-\ell}^{+\ell} \left[{}_{s'} Y^*_{\ell,m}(\theta',\phi')\right] \left[{}_s Y_{\ell,m}(\theta,\phi)\right] & = & \left(\frac{2\ell+1}{4\pi}\right) \sum_{m=-\ell}^{+\ell} D^{\ell *}_{-s',m}(\phi',\theta',0) D^{\ell}_{-s,m}(\phi,\theta,0) \\
   & = & \left(\frac{2\ell+1}{4\pi}\right) D^{\ell}_{-s,-s'}(\alpha,\beta,\gamma) \\
   & = & \sqrt{\frac{2\ell+1}{4\pi}} \text{e}^{-i s \gamma} {}_{s} Y_{\ell,-s'}(\beta,\alpha) \\
\end{eqnarray}
with $(\alpha,\beta,\gamma)$ the Euler angles of the rotation bringing the direction $(\theta',\phi')$ to $(\theta,\phi)$.

\section{Computing the Wigner $d$ function}
\label{app:wignerd}

We use the decomposition of the Wigner $d$ function in terms of their Fourier representation. This decomposition is taken from \cite{Edmonds57}. We start by factorizing a nodal rotation:
\begin{equation}
  R(0,\beta,0) = R\left(-\frac{\pi}{2},0,0\right) R\left(0,-\frac{\pi}{2},0\right) R\left(\beta,0,0\right) R\left(0,\frac{\pi}{2},0\right) R\left(\frac{\pi}{2},0,0\right).
\end{equation}
Expressing this matrix multiplication in terms of the elements of the $D$ matrices yields the identity:
\begin{equation}
  d^\ell_{m_1,m_2}(\beta) = \imag^{m_2-m_1} \sum_{n=-\ell}^\ell d^\ell_{n,m_1}\left(\frac{\pi}{2}\right) d^\ell_{n,m_2}\left(\frac{\pi}{2}\right) \mathrm{e}^{\imag n \beta} \label{eq:fft_dfunction}.
\end{equation}
If we let $\Delta^\ell_{m_1,m_2} = d^\ell_{m_1,m_2}\left(\frac{\pi}{2}\right)$ and $B^\ell_{n,m_1,m_2}=\imag^{m_1-m_2} \Delta^\ell_{n,m_1} \Delta^\ell_{n,m_2}$, then we see explicitly the expression of $d^\ell_{m_1,m_2}(\beta)$ in terms of a discrete Fourier transform:
\begin{equation}
  d^\ell_{m_1,m_2}(\beta) = \sum_{n=-\ell}^\ell B^\ell_{n,m_1,m_2} \mathrm{e}^{\imag n \beta}.
\end{equation}

The recursive formula for $d$-matrix can be adapted for the specific case of $\beta=\pi/2$. This calculation yields the following recursion formula:
\begin{gather}
  \Delta^\ell_{\ell,0} = -\left(\frac{2 \ell - 1}{2 \ell}\right)^{1/2} \Delta^{\ell-1}_{\ell-1,0} \label{eq:delta_lorder} \\
  \Delta^\ell_{\ell,m_2} = \left[ \frac{(\ell/2)(2 \ell - 1)}{(\ell+m_2)(\ell+m_2-1)} \right]^{1/2} \label{eq:delta_1col} \Delta^{\ell-1}_{\ell-1,m_2-1} \\
  \begin{split}
    \Delta^\ell_{m_1,m_2} = 
      \frac{2 m_2}{[(\ell-m_1)(\ell+m_1+1)]^{1/2}} \Delta^\ell_{m_1+1,m_2} \\
      - \left[\frac{(\ell-m_1-1)(\ell+m_1+2)}{(\ell-m_1)(\ell+m_1+1} \right]^{1/2} \Delta^\ell_{m_1+2,m_2}
  \end{split} \label{eq:delta_2col}
\end{gather}
with an initial condition
\begin{equation}
  \Delta^0_{0,0} = 1.
\end{equation}
We start by using Eq.~\eqref{eq:delta_lorder} to increase the order in
$\ell$ from the initial condition. At the same time, we compute all
$\Delta^\ell_{\ell,m}$ for $0\le m \le \ell$ using Eq.~\eqref{eq:delta_1col}
recursively. At the end of the recursion we have access to all
$\Delta^\ell_{\ell,m}$ at the required $\ell$. We use the last equation
\eqref{eq:delta_2col} to compute $\Delta^\ell_{u,m_1}$ and
$\Delta^\ell_{u,m_2}$ for all $-\ell \le u \le \ell$ at $m_1$ and $m_2$ fixed.

The function $d^\ell_{m_1,m_2}(\beta)$ is used to compute the angular
correlation function described in Eq.~\eqref{eq:spins_correl}. We
decided to tabulate and interpolate using cubics the correlation
function. For performance reason, we compute the Wigner-$d$ function at all
$\beta$ at sufficiently high resolution and then sum the contribution
at each $\beta$ for any given $\ell$ of the whole summation of
$d^\ell_{-s,-s}$.

\end{document}